\documentclass[sigconf]{acmart}

\acmConference[ISSTA 2024]{ACM SIGSOFT International Symposium on Software Testing and Analysis}{16-20 September, 2024}{Vienna, Austria}
\AtBeginDocument{%
  }

\setcopyright{acmcopyright}
\copyrightyear{2018}
\acmYear{2018}
\acmDOI{XXXXXXX.XXXXXXX}

\acmPrice{15.00}
\acmISBN{978-1-4503-XXXX-X/18/06}

\usepackage{multirow}
\usepackage{subcaption}
\usepackage{utfsym}
\usepackage{tikz}
\newcommand{\circled}[1]{\lower.7ex\hbox{\tikz\draw (0pt, 0pt)
circle (.5em) node {\makebox[1em][c]{\small #1}};}}

\begin{document}

\newcommand{\MethodName}{iJTyper}

\title{iJTyper: An Iterative Type Inference Framework for Java by Integrating Constraint- and Statistically-based Methods}



\author{Zhixiang Chen}
\affiliation{%
  \department{School of Software Engineering}
  \institution{Sun Yat-sen University}
  \country{China}}
\email{chenzhx69@mail2.sysu.edu.cn}

\author{Anji Li}
\affiliation{%
  \department{School of Software Engineering}
  \institution{Sun Yat-sen University}
  \country{China}}
\email{lianj8@mail2.sysu.edu.cn}

\author{Neng Zhang*}
\affiliation{%
  \department{School of Software Engineering}
  \institution{Sun Yat-sen University}
  \country{China}}
\email{zhangn279@mail.sysu.edu.cn}

\author{Jianguo Chen}
\affiliation{%
  \department{School of Software Engineering}
  \institution{Sun Yat-sen University}
  \country{China}}
\email{chenjg33@mail.sysu.edu.cn}

\author{Yuan Huang}
\affiliation{%
  \department{School of Software Engineering}
  \institution{Sun Yat-sen University}
  \country{China}}
\email{huangyuan5@mail.sysu.edu.cn}

\author{Zibin Zheng}
\affiliation{%
  \department{School of Software Engineering}
  \institution{Sun Yat-sen University}
  \country{China}}
\email{zhzibin@mail.sysu.edu.cn}

\renewcommand{\shortauthors}{Trovato et al.}

\begin{abstract}
  Inferring the types of API elements in incomplete code snippets (e.g., those on Q\&A forums) is a prepositive step required to work with the code snippets. Existing type inference methods can be mainly categorized as constraint-based or statistically-based. The former imposes higher requirements on code syntax and often suffers from low recall due to the syntactic limitation of code snippets. The latter relies on the statistical regularities learned from a training corpus and does not take full advantage of the type constraints in code snippets, which may lead to low precision. 
In this paper, we propose an iterative type inference framework for Java, called \MethodName{}, by integrating the strengths of both constraint- and statistically-based methods. For a code snippet, \MethodName{} first applies a constraint-based method and augments the code context with the inferred types of API elements. \MethodName{} then applies a statistically-based method to the augmented code snippet. The predicted candidate types of API elements are further used to improve the constraint-based method by reducing its pre-built knowledge base. \MethodName{} iteratively executes both methods and performs code context augmentation and knowledge base reduction until a termination condition is satisfied. Finally, the final inference results are obtained by combining the results of both methods. We evaluated \MethodName{} on two open-source datasets. Results show that 1) \MethodName{} achieves high average precision/recall of 97.31\% and 92.52\% on both datasets; 2) \MethodName{} significantly improves the recall of two state-of-the-art baselines, SnR and MLMTyper, by at least 7.31\% and 27.44\%, respectively; and 3) \MethodName{} improves the average precision/recall of the popular language model, ChatGPT, by 3.25\% and 0.51\% on both datasets.
\end{abstract}

\ccsdesc[500]{Software and its engineering}
\ccsdesc[100]{Maintaining software~Code Analysis}

\keywords{Type Inference, Java, Method Integration, Iterative algorithm}


\maketitle

\section{Introduction}
Code snippets are widely used in online Q\&A forums such as Stack Overflow (SO) and GitHub Issues to illustrate problems or provide examples of API usages~\cite{sadowski2015developers}. 
Software developers often search for code snippets 
to find solutions to coding problems \cite{bacchelli2012harnessing,barua2014developers,rosen2016mobile}, which can save valuable time and effort from learning extensive knowledge they have not yet grasped. 
However, online code snippets often lack completeness 
as they do not provide sufficient dependency information to determine the exact types of API elements from external libraries~\cite{DBLP:conf/issta/TerragniLC16-csnippex, dagenais2012recovering}. 
As a result, the code snippets generally cannot be compiled or reused directly~\cite{DBLP:conf/msr/YangHL16,horton2018gistable}. According to \cite{DBLP:conf/issta/TerragniLC16-csnippex}, more than 91\% code snippets on SO cannot be compiled due to various errors, among which the missing type declarations of API elements is the most common error that accounts for 38\%. Therefore, inferring the types of API elements in code snippets is a prepositive step required to work with those code snippets~\cite{DBLP:conf/kbse/HuangYXX0022-prompt,DBLP:conf/icse/DongGTS22,DBLP:conf/kbse/SaifullahAR19-coster}.

Existing type inference methods for incomplete code snippets can be primarily categorized as constraint-based~\cite{DBLP:conf/icse/DongGTS22,DBLP:journals/corr/abs-2108-01165-Depres,DBLP:conf/icse/SubramanianIH14-baker} or statistically-based~\cite{DBLP:conf/icse/PhanNTTNN18-statype,DBLP:journals/scp/VelazquezRodriguezNR23-resico,DBLP:conf/kbse/SaifullahAR19-coster,DBLP:conf/kbse/HuangYXX0022-prompt}. Constraint-based methods often pre-build a knowledge base from API libraries and extract type constraints from code snippets. Then, they employ heuristic rules to find constraint-satisfying solutions. Although these methods could achieve high precision, they impose higher requirements on the code syntax, which often results in low recall because of the syntactic limitation of code snippets. Statistically-based methods build statistical models by learning from extensive code corpus and predict types with the greatest likelihood in the context of API elements. They are hardly affected by syntax limitation and thus could achieve relatively high recall. However, these methods generally 
do not take advantage of the type constraints in code snippets, which may lead to low precision. \textbf{The complementary nature possessed by the two kinds of methods makes their combination rational, which forms the basis of this work.}

In this paper, we propose \MethodName{}, an iterative type inference framework for Java code snippets by integrating constraint- and statistically-based methods. Given a code snippet, \MethodName{} first applies a constraint-based method, e.g., SnR~\cite{DBLP:conf/icse/DongGTS22}, to it. The inferred types of API elements are used to augment the context of the code snippet. Then, the augmented code snippet is fed to a statistically-based method, e.g., MLMTyper~\cite{DBLP:conf/kbse/HuangYXX0022-prompt}. The candidate types predicted for API elements are further used to reduce irrelevant types in the pre-built knowledge base (KB) of the constraint-based method in order to improve the performance of the method. \MethodName{} iteratively performs both methods as well as the code context augmentation and KB reduction until a termination condition (e.g., the inference results of both methods become stable) is reached. \MethodName{} produces the final inference results by combining the results of both methods. 

To evaluate \MethodName{}, we conducted a series of experiments on two open-source datasets 
collected from SO: StatType-SO~\cite{DBLP:conf/icse/PhanNTTNN18-statype} and~Short-SO~\cite{DBLP:conf/kbse/HuangYXX0022-prompt}. Two state-of-the-art (SOTA) type inference methods, 
SnR and MLMTyper, were selected as the baselines. Moreover, we compared \MethodName{} with the popular language model, ChatGPT~\cite{ChatGPT}. The results are as follows. 
\MethodName{} achieves the highest average precision/recall of 97.31\% and 92.52\% on StatType-SO and Short-SO, significantly outperforming SnR and MLMTyper by at least 7.31\% and 27.44\%, respectively. \MethodName{} improves the average precision/recall of ChatGPT by 3.25\% on StatType-SO and by 0.51\% on Short-SO. 
In summary, the two main contributions of this paper are listed below:
\begin{itemize}
    \item We propose \MethodName{}, an iterative type inference framework for Java by integrating the strengths of constraint- and statistically-based type inference methods. To the best of our knowledge, we are the first to integrate the two kinds of methods. In addition, \MethodName{} is flexible and can be easily adapted to integrate different methods.
    \item We conducted extensive experiments to evaluate \MethodName{}, in comparison with two SOTA baselines and ChatGPT. The results demonstrate the superiority of \MethodName{}. 
\end{itemize}

\section{Motivation}
\subsection{Type Inference Task Description}\label{subsec:titd}
  
\begin{figure}
    \centering
    \includegraphics[width=\linewidth]{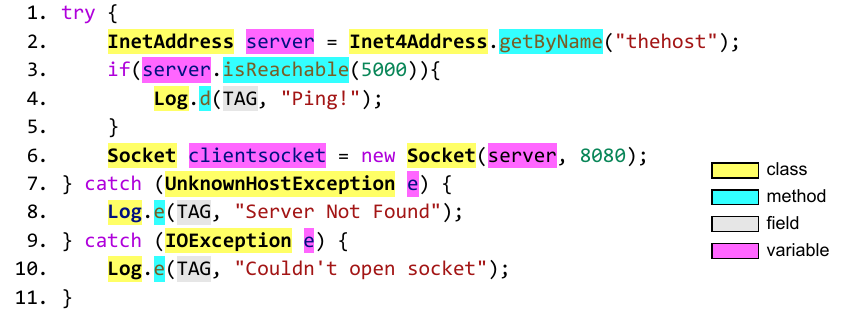}
    \caption{Example code snippet annotated with four kinds of API elements from the SO post `442496'.}
    \label{fig:problem-description}
\end{figure}

The type inference task aims to identify the types (i.e., fully qualified names (FQNs)) of API elements in an incomplete code snippet. In general, there are five kinds of API elements whose types need to be inferred, i.e., \emph{classes}, \emph{interfaces}, \emph{methods}, \emph{fields}, and \emph{variables}.
Fig.~\ref{fig:problem-description} shows an example code snippet from the SO post `442496'. The classes, methods, fields, and variables are marked with yellow, blue, grey, and pink backgrounds, respectively. 

Among the API elements, classes and interfaces are more important since their types, together with the syntax knowledge (e.g., the declared methods and fields and the inheritance and implementation relationships) of them, can be used to infer the types of the other elements. For example, for the code snippet shown in Fig.~\ref{fig:problem-description}, after determining the type of the class `\emph{Inet4Address}' as `\emph{java.net.Inet4Address}', the type of the method `\emph{getByName("thehost")}' invoked by the class can be inferred as `\emph{public static InetAddress getByName(String host) throws UnknownHostException}' by leveraging the syntax knowledge of the class\footnote{https://docs.oracle.com/en/java/javase/17/docs/api/java.base/java/net/Inet4Address.html}. The type of the variable \emph{server} can be further inferred based on the return type of the method. Based on this insight, existing type inference methods primarily focus on identifying the types of classes and interfaces. Moreover, the classes that correspond to primitive data types, e.g., \emph{int} and \emph{Integer}, are often excluded because such types are easy to recognize and thus inferring them is meaningless. \textbf{Similarly, in this work, we focus on the type inference of classes and interfaces except those corresponding to primitive data types when describing our proposed \MethodName{} framework.}

\subsection{Limitations of Type Inference Methods}\label{sec:mlim}

Existing type inference methods proposed for Java code snippets can be categorized into two groups: constraint-based~\cite{DBLP:conf/icse/SubramanianIH14-baker,DBLP:conf/kbse/Shokri21-ase_depres,DBLP:conf/icse/DongGTS22} and statistically-based~\cite{DBLP:conf/icse/PhanNTTNN18-statype,DBLP:conf/kbse/SaifullahAR19-coster,DBLP:journals/scp/VelazquezRodriguezNR23-resico,DBLP:conf/kbse/HuangYXX0022-prompt}.

\begin{table}
    \centering
    \caption{Constraint-based type inference methods.}
    \label{tab:constraint-methods-review}
    \resizebox{\linewidth}{!}
    {
        \begin{tabular}{l|c|l} 
            \hline
            \multicolumn{1}{c|}{\textbf{Method}} & \textbf{Publication} & \multicolumn{1}{c}{\textbf{Summary}}                                                                                                                                                                                     \\ 
            \hline
            Baker~\cite{DBLP:conf/icse/SubramanianIH14-baker}                                & ICSE 2014            & \begin{tabular}[c]{@{}l@{}}Baker maintains a candidate list for each API and applies heuristic~rules to reduce the\\number of candidates deductively.\end{tabular}                                                       \\ 
            \hline
            DepRes~\cite{DBLP:conf/kbse/Shokri21-ase_depres}                               & ASE 2021             & \begin{tabular}[c]{@{}l@{}}DepRes generates a sketch for each API and utilizes a Z3 SMT~Solver to handle the \\constraints, aiming to find a minimum number~of types and dependencies.\end{tabular}                      \\ 
            \hline
            SnR~\cite{DBLP:conf/icse/DongGTS22}                                  & ICSE 2022            & \begin{tabular}[c]{@{}l@{}}SnR builds a knowledge base for generics and utilizes Datalog~to solve the~constraints. \\It ranks the candidates based on a~prioritization heuristic and selects the top item.\end{tabular}  \\ 
            \hline
            \multicolumn{1}{l}{}                 & \multicolumn{1}{l}{} &                                                                                                                                                                                                                         
            \end{tabular}
    }
\end{table}

\subsubsection{Constraint-based Type Inference Methods.}\label{subsec:ctim} 
A constraint-based method typically pre-builds a KB that stores the syntax knowledge of API elements from a set of API libraries. Given a code snippet, it extracts type constraints between API elements hidden in the code snippet, e.g., the methods invoked by an object variable of a class. Finally, the types of API elements that can satisfy the constraints at most are determined based on the KB.

Table~\ref{tab:constraint-methods-review} presents a list of constraint-based methods for Java. SnR achieves the best performance according to the evaluations reported in~\cite{DBLP:conf/icse/DongGTS22, DBLP:conf/kbse/SaifullahAR19-coster}. 
SnR pre-builds a KB from a set of Jar files by extracting the classes and interfaces, their included methods and fields, and the inheritance and implementation relationships between them. For an input code snippet, SnR attempts to repair it to a syntactically valid compilation unit using a template-based method, such that an Abstract Syntax Tree (AST) can be generated for the unit. After that, SnR extracts type constraints from the AST and represents the constraints using a declarative logic programming language, Datalog~\cite{DBLP:journals/pacmpl/MadsenL20-datalog}. Based on the KB and constraints, several heuristic rules are used for type inference, e.g., minimizing the number of unique libraries used in the code snippet.

\begin{figure}
    \centering
    \includegraphics[width=\linewidth]{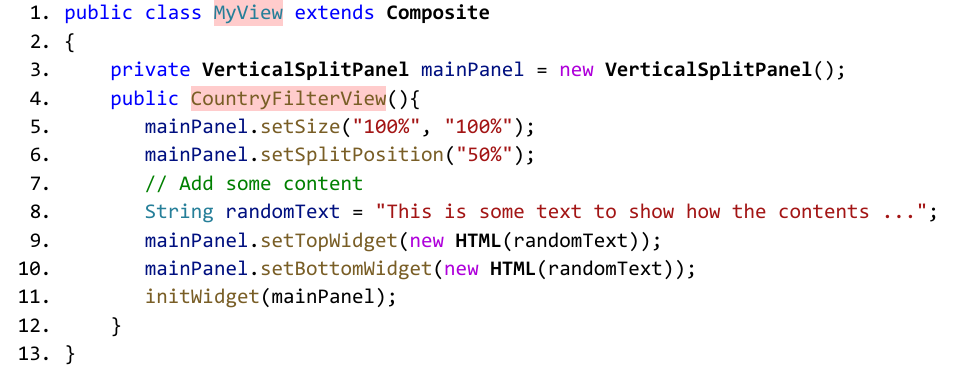}
    \caption{Example code snippet with syntax errors from the SO post `1318732'. A syntax error is that the constructor name, \emph{CountryFilterView}, is not the same as the class name, \emph{MyView}.}
    \label{fig:syntax-incorrectness}
\end{figure}

A major limitation of constraint-based methods is that they require the valid compilation of the input code snippet in order to extract constraints for type inference. Although some methods, e.g., SnR, address this problem using a template-based syntax repair method, there still exists a large portion of code snippets that cannot be repaired or can only be partially repaired, which obstructs the application of constraint-based methods. For example, 
the syntax repair method used by SnR only focuses on 38\% of the total compilation errors in code snippets from SO posts, and it can only address 16.7\% of the focused errors~\cite{DBLP:conf/issta/TerragniLC16-csnippex}. Fig.~\ref{fig:syntax-incorrectness} shows a code snippet from the SO post `1318732'. Table~\ref{tab:snr-syntax-incorrectness} presents the type inference results produced by SnR. SnR only infers the type of the class \emph{Composite}[1,1] (the two indices in `[]' are explained in the caption of the table) outside the definition of the class \emph{MyView}
and fails to infer the types of other classes. This issue is caused by the syntax error that the constructor name, \emph{CountryFilterView}, is different from the class name, \emph{MyView}.

\begin{figure}[t]
    \centering
    \includegraphics[width=\linewidth]{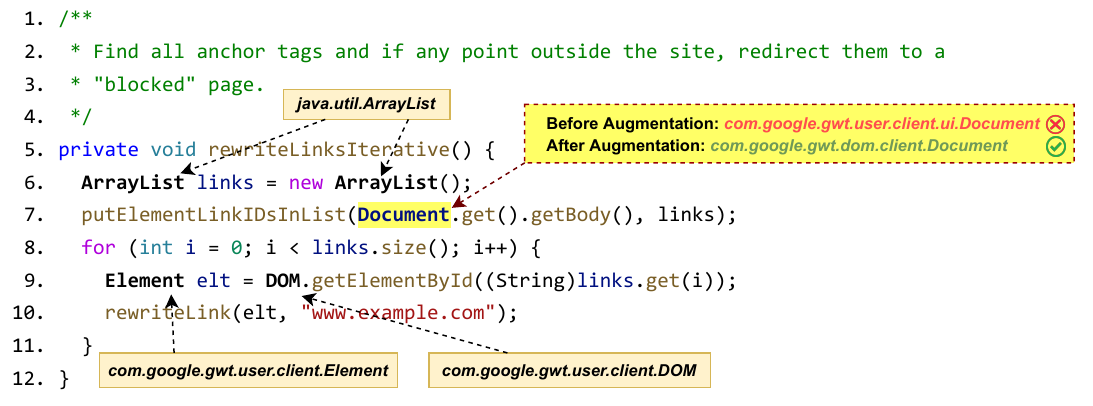}
    \caption{Example of the type inference results improved by leveraging the augmented code snippet using a code snippet from the SO post `3954392'. The types with a brown background are inferred by applying SnR to the original code snippet. The types in the yellow background box are predicted by MLMTyper before and after augmenting the original code snippet with the inferred types of SnR.}
    \label{fig:context-augment}
\end{figure}

Apart from the compilation limitation, constraint-based methods are also limited by 1) the coverage of the pre-built KB; 2) the richness of the type constraints included in the code snippet; and 3) the capability of the methods in extracting the constraints. Specifically, the types of API elements outside the KB cannot be inferred. For a code snippet, if an API element has multiple candidate types in the KB, in order to determine the type of the element, the code snippet should contain sufficient constraints for filtering the other candidates; and meanwhile the methods should be able to extract the constraints and use them for type inference effectively.
Fig.~\ref{fig:context-augment} shows a code snippet from the SO post `3954392', which 
illustrates the limitation of SnR on the capability of extracting constraints from code snippets. By applying SnR to it, four classes get their right types, 
but no type is inferred for \emph{Document}[7,1]. Through an in-depth analysis, we find that the class has nine candidate types in the pre-built KB. Although the KB contains the right type, \emph{com.google.gwt.dom.client.Document}, SnR fails to determine the type. This issue is because SnR cannot extract the full constraints expressed by the cascaded method calls, \emph{Document.get().getBody()}, and thus cannot distinguish the right type from a noise type, \emph{com.extjs.gxt.ui.client.widget.Document}, which also has a method \emph{get()} but the return type does not have a method \emph{getBody()}.

\begin{table}
    \centering
    \caption{Type inference results produced by two SOTA methods, SnR and MLMTyper, for the code snippet shown in Fig.~\ref{fig:syntax-incorrectness}. To distinguish multiple API elements with the same name in the code snippet, two indices (i.e., the line number and the order of occurrence) are used to locate a specific API element, e.g., `VerticalSplitPanel$[\underline{3},\underline{2}]$' indicates the \underline{2nd} class `VerticalSplitPanel' at \underline{line 3}.}
    \label{tab:snr-syntax-incorrectness}
        \resizebox{\linewidth}{!}
        {
            \begin{tabular}{l|c|l|c|l} 
                \toprule
                \multicolumn{1}{c|}{\textbf{API}} & \multicolumn{2}{c|}{\textbf{\textbf{SnR}}}                       & \multicolumn{2}{c}{\textbf{MLMTyper}}                                      \\ 
                \hline
                Composite[1,1]                    & \textit{com.google.gwt.user.client.ui.Composite} & $\usym{2714}$ & \textit{android.widget.Composite}                         & $\usym{2717}$  \\
                VerticalSplitPanel[3,1]           & -                                                &               & \textit{com.google.gwt.user.client.ui.VerticalSplitPanel} & $\usym{2714}$  \\
                VerticalSplitPanel[3,2]           & -                                                &               & \textit{com.google.gwt.user.client.ui.VerticalSplitPanel} & $\usym{2714}$  \\
                HTML[9,1]                         & -                                                &               & \textit{com.google.gwt.user.client.ui.HTML}               & $\usym{2714}$  \\
                HTML[10,1]                        & -                                                &               & \textit{com.google.gwt.user.client.ui.HTML}               & $\usym{2714}$  \\
                \bottomrule
                \end{tabular}
        }
\end{table}

\subsubsection{Statistically-based Type Inference Methods.} A statistically-based method builds a statistical model from an extensive code corpus using machine learning (including deep learning) techniques. The types of API elements in the code corpus are given. Using the model, the method infers the types of API elements in a code snippet by maximizing the likelihood of the code snippet. Statistically-based methods eliminate the compilation limitation of code snippets by treating them as plain text.

The common basis of statistically-based methods is the software naturalness~\cite{DBLP:journals/cacm/HindleBGS16-naturalness}. That is, human-written code is mostly simple, repetitive, and predictable, and thus can be modeled by statistical language models to facilitate various software engineering tasks, e.g., type inference~\cite{DBLP:conf/kbse/HuangYXX0022-prompt,DBLP:journals/scp/VelazquezRodriguezNR23-resico,DBLP:conf/icse/PhanNTTNN18-statype,DBLP:conf/kbse/SaifullahAR19-coster}, defect localization~\cite{qiu2021deep-defect-localization,zhang2020softwareattention-defect-localization}, and code completion~\cite{raychev2014statistical-codecompletion,svyatkovskiy2019pythia-code-completion}. Code context is used to capture such naturalness~\cite{DBLP:conf/kbse/SaifullahAR19-coster}. The key difference between statistically-based methods is the code context that they employ to predict the types of API elements. Table~\ref{tab:statistical-based-review} presents a list of statistically-based methods for Java. For example, STATTYPE~\cite{DBLP:conf/icse/PhanNTTNN18-statype} leverages type context and resolution context. The former refers to the classes, methods, and fields that occur around an API element, $e$, to be inferred, while the latter refers to the type inference decision of the API elements surrounding $e$. The prompt-tuned masked language model (MLM)-based method proposed by Huang et al.~\cite{DBLP:conf/kbse/HuangYXX0022-prompt}, referred to as MLMTyper, uses the code line where $e$ is located and the top and down $\eta$ (=2 by default) adjacent code lines to form a code context block.

Based on the evaluations reported in~\cite{DBLP:conf/kbse/HuangYXX0022-prompt,DBLP:journals/scp/VelazquezRodriguezNR23-resico}, 
MLMTyper achieves the best performance. It transforms the type inference task into a cloze-filling problem and uses a prompt-tuned MLM to complete the missing types. MLMTyper achieves a relatively high recall of 91\%, which is better than that (i.e., 87.5\%) of SnR. The improvement is mainly because MLMTyper overcomes the compilation limitation and would not fail to give inference in any cases.

\begin{table}
    \centering
        \caption{Statistically-based type inference methods.}
        \label{tab:statistical-based-review}
        \resizebox{\linewidth}{!}
        {
            \begin{tabular}{l|c|l} 
                \hline
                \multicolumn{1}{c|}{\textbf{Method}} & \textbf{Publication} & \multicolumn{1}{c}{\textbf{Summary}}                                                                                                                                                                                                                                                                   \\ 
                \hline
                STATTYPE~\cite{DBLP:conf/icse/PhanNTTNN18-statype}                             & ICSE 2018            & \begin{tabular}[c]{@{}l@{}}Based on the assumption that APIs have co-occurring regularities which can be learned\\by statistical models, STATTYPE treats the type inference task as a statistical machine \\translation~task and considers both the type context and resolution context.\end{tabular}  \\ 
                \hline
                COSTER~\cite{DBLP:conf/kbse/SaifullahAR19-coster}                               & ASE 2019             & \begin{tabular}[c]{@{}l@{}}COSTER utilizes both the local and global contexts of an API. It calculates a likelihood score, \\a~context similarity score, and a name similarity score based on a pre-built occurrence \\likelihood~dictionary to map an API element to its type.\end{tabular}           \\ 
                \hline
                RESICO~\cite{DBLP:journals/scp/VelazquezRodriguezNR23-resico}                               & SCP 2023             & \begin{tabular}[c]{@{}l@{}}RESICO uses the Word2Vec model to vectorize the API reference and its context, which \\is then fed into a machine learning classifier to obtain type recommendations.\end{tabular}                                                                                          \\ 
                \hline
                MLMTyper~\cite{DBLP:conf/kbse/HuangYXX0022-prompt}                             & ASE 2022             & \begin{tabular}[c]{@{}l@{}}MLM transforms the type inference task into a fill-in-blank task based on a prompt-tuned \\masked language model.\end{tabular}                                                                                                                                              \\
                \hline
                \end{tabular}
        }
\end{table}

\begin{figure}
    \centering
    \includegraphics[width=\linewidth]{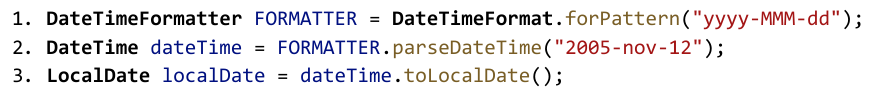}
    \caption{Example code snippet from the SO post `8746084'.}
    \label{fig:ase-disadvantage}
\end{figure}

A major limitation of statistically-based methods is that they rarely take the syntax knowledge of APIs and the type constraints in code snippets into consideration, which may lead to low precision. This limitation becomes more obvious when facing APIs or contexts that are less frequent in the training code corpus~\cite{DBLP:conf/kbse/HuangYXX0022-prompt, DBLP:conf/kbse/SaifullahAR19-coster}. Fig.~\ref{fig:ase-disadvantage} shows a code snippet from the SO post `8746084'. Table~\ref{tab:MLM-disadvantage-example} presents the top-1 types of the classes inferred by MLMTyper. 
MLMTyper correctly predicts \emph{DateTime}[2,1] as \emph{org.joda.time.DateTime} but mistakenly predicts the types of the other three classes. For example, the return type of the method \emph{org.joda.time.DateTime.toLocalDate()} is \emph{org.joda.time.LocalDate}. However, MLMTyper does not consider the syntax knowledge and predicts the wrong type for \emph{LocalDate}[3,1].

\subsubsection{Insight.}
As explained above, both constraint- and statistically-based type inference methods have their own limitations. Constraint-based methods often have relatively high precision by leveraging the syntax knowledge of APIs and the type constraints in code snippets, however their recall may be low due to several factors, e.g., the compilation failure of code snippets and the low coverage of the pre-built KB. In contrast, statistically-based methods do not require to compile code snippets and can infer types for all API elements, which could improve the recall. However, their precision may be low since they do not consider the constraints in code snippets and thus may lead to wrong types. Tables~\ref{tab:snr-syntax-incorrectness} and~\ref{tab:MLM-disadvantage-example} present the type inference results produced by SnR and MLMTyper for two code snippets. SnR achieves a high precision of 100\% in both cases but a low recall of 20\% in the first case. Compared with SnR, MLMTyper achieves a higher recall of 80\% and also a lower precision of 80\% in the first case. In the second case, both the precision and recall of MLMTyper are 25\% and are much lower than those of SnR.

\begin{table}
    \centering
    \caption{Type inference results produced by SnR and MLMTyper for the code snippet shown in Fig.~\ref{fig:ase-disadvantage}. Similar to Table~\ref{tab:snr-syntax-incorrectness}, two indices (i.e., the line number and the order of occurrence) are used to locate a specific API element in the code snippet.}
    \label{tab:MLM-disadvantage-example}
    \resizebox{\linewidth}{!}{
        \begin{tabular}{l|c|l|c|l} 
            \toprule
            \multicolumn{1}{c|}{\textbf{API}} & \multicolumn{2}{c|}{\textbf{SnR}}                               & \multicolumn{2}{c}{\textbf{MLMTyper}}                 \\ 
            \hline
            DateTimeFormatter[1,1]            & \textit{org.joda.time.format.DateTimeFormatter} & $\usym{2714}$ & \textit{java.text.DateTimeFormatter} & $\usym{2717}$  \\
            DateTimeFormat[1,1]               & \textit{org.joda.time.format.DateTimeFormat}    & $\usym{2714}$ & \textit{java.text.DateTimeFormat}    & $\usym{2717}$  \\
            DateTime[2,1]                     & \textit{org.joda.time.DateTime}                 & $\usym{2714}$ & \textit{org.joda.time.DateTime}      & $\usym{2714}$  \\
            LocalDate[3,1]                    & \textit{org.joda.time.LocalDate}                & $\usym{2714}$ & \textit{java.time.LocalDate}         & $\usym{2717}$  \\
            \bottomrule
            \end{tabular}
    }
\end{table}

\textbf{We further observe from the table results that the type inference results produced by SnR and MLMTyper exhibit a certain degree of complementarity.} As an example, for the code snippet shown in Fig.~\ref{fig:syntax-incorrectness}, SnR correctly infers the type of the class \emph{Composite}[1,1] but fails to infer the types of the other classes, while MLMTyper predicts wrong type for \emph{Composite}[1,1] but correct types for the others. Based on this observation, better performance could be achieved by integrating constraint- and statistically-based methods. Specifically, the limitations of constraint-based methods can be partially addressed by statistically-based methods because they have two main advantages: 1) They are not affected by the syntax errors in code snippets; and 2) They employ the context (e.g., the token co-occurrence) of API elements to predict the types without the need to extract constraints from code snippets. 
Moreover, the types inferred by the constraint-based method could be used to augment the context of API elements in the code snippet and thus improve the performance of the statistically-based method. Following this insight, we propose an integrated type inference framework, which is elaborated in the following section.

\section{Methodology}\label{sec:meth}
\begin{figure}[t]
    \centering
    \includegraphics[width=\linewidth]{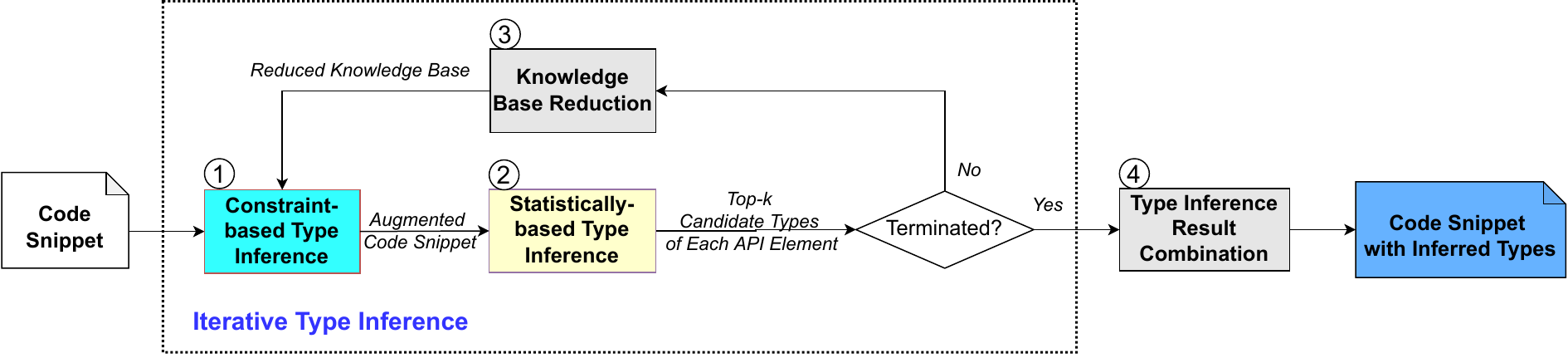}
    \caption{The framework of \MethodName.}
    \label{fig:overview}
\end{figure}

Fig.~\ref{fig:overview} shows the framework of \MethodName{}. The input is a code snippet, and the output is the code snippet with inferred types of the API elements. 
\MethodName{} contains four steps \circled{1}-\circled{4}. 
At first, \MethodName{} applies a constraint-based type inference method, e.g., SnR~\cite{DBLP:conf/icse/DongGTS22}, to the code snippet (\circled{1}). Since the inference results may contain missing or wrong types due to the limitations of the method (see Section~\ref{subsec:ctim}), 
\MethodName{} further feeds the code snippet augmented with the inferred types to a statistically-based type inference method, e.g., MLMTyper~\cite{DBLP:conf/kbse/HuangYXX0022-prompt} (\circled{2}). In turn, the top-$k$ candidate types predicted by the statistically-based method could be used to improve the constraint-based method by filtering irrelevant types in the pre-built KB. Thus, \MethodName{} adopts an iterative mechanism. After performing a round of both methods, \MethodName{} examines whether a termination condition (e.g., the inference results become stable) is satisfied. If the answer is negative, \MethodName{} reduces the KB of the constraint-based method using the top-$k$ candidate types of each API element predicted by the statistically-based method (\circled{3}) and then re-applies the constraint-based method to the code snippet. After the iteration is terminated, \MethodName{} combines the inference results of both methods to produce the final inference results (\circled{4}).

Notice that \MethodName{} is a flexible framework that can integrate different constraint- and statistically-based methods. In this work, we demonstrate the framework using two SOTA methods, SnR and MLMTyper, for instance. In addition, based on our literature review, some constraint-based methods are complex and their replication packages are difficult to understand and modify. For example, the implementation code of SnR is obfuscated and compiled to a Jar file, making it hard to replicate and extend the work. 
We provide a simple solution to integrating any constraint-based method which has a workable replication package. Instead of adjusting the implementation code, \MethodName{} directly reduces the pre-built KB based on the top-$k$ candidate types predicted by the statistically-based method. \textbf{We acknowledge that this solution may sacrifice some efficiency, however it can help users escape from the efforts required to understand and modify the implementation of the constraint-based method and thus could be desired by users in non real-time application scenarios.}

\subsection{Constraint-based Type Inference}

For an input code snippet, we first use a constraint-based type inference method to infer the types of API elements. As explained previously,  due to the limitations of the method (see Section~\ref{subsec:ctim}), the inference may fail or the inferred types may be incorrect for some API elements, as demonstrated in Figs.~\ref{fig:syntax-incorrectness} and~\ref{fig:context-augment}. 
However, since the precision of the constraint-based method is relatively high, the inferred types could be used to enhance the context of API elements in the code snippet and thus improve the statistically-based method. We build the augmented code snippet by replacing the API elements with their inferred types, which is formalized as follows.

\textbf{Context Augmentation of the Code Snippet.}
Recall that statistically-based methods treat code snippets as plain text. Specifically, a code snippet can be represented as a token sequence, $CS=t_{1}t_{2}...t_{m}$ where some of the tokens are API elements. The set of API elements, denoted as $APIs(CS)$, can be identified from the AST of the code snippet or using regular expressions if the AST generation fails due to syntax errors. Based on the type inference results of the constraint-based method, $APIs(CS)$ can be divided into two subsets: the set of API elements with inferred types and the set of API elements without inferred types, which are denoted as $Typed\_APIs^{C}(CS)$ and $NonTyped\_APIs^{C}(CS)$, respectively. The superscript `$C$' stands for the constraint-based method. The inferred type of an API element $t_{i}\in Typed\_APIs^{C}(CS)$ is denoted as $type^{C}(t_{i})$. Then, the augmented code snippet can be represented as a new token sequence, $ACS=y_{1}y_{2}...y_{m}$ where the API elements are replaced with their types, i.e., 
\begin{equation}
    y_{i} = type^{C}(t_{i}) \text{ if } t_{i} \in Typed\_APIs^{C}(CS) \text{ else } t_{i}, \forall i = 1,...,m   
\end{equation}

\subsection{Statistically-based Type Inference}

In this step, a statistically-based type inference method is applied to the augmented code snippet $ACS$. 

We perform this step using the statistically-based method, MLMTyper. MLMTyper has two execution modes which are called \emph{leave-one-out} and \emph{all-unknown} in~\cite{DBLP:conf/kbse/HuangYXX0022-prompt}. Specifically, MLMTyper predicts each API element in a code snippet one-by-one. For a specific API element, $e$, after building the code context block, the leave-one-out mode predicts the top-$k$ candidate types of $e$ with the known types of all the other API elements in the context. In contrast, the all-unknown mode predicts the types of $e$ without utilizing the types of any other API elements, i.e., assuming that those types are all unknown. The leave-one-out and all-unknown modes respectively represent the upper bound (i.e., the easiest case) and lower bound (i.e., the most difficult case) of type inference scenarios. We execute MLMTyper with the leave-one-out mode on $ACS$. Note that our execution is not strictly a leave-one-out mode because there can be some missing and/or incorrect types in $ACS$. 
Considering the relatively high precision of SnR (see Table~\ref{tab:rq2-component}), the majority of the types in $ACS$ should be correct and thus can improve MLMTyper in most cases. Taking the code snippet shown in Fig.~\ref{fig:context-augment} as an example, before augmenting it, the top-1 type of the class \emph{Document}[7,1] predicted by MLMTyper, i.e., \emph{com.google.gwt.user.client.ui.Document}, is wrong, which is haphazardly generated by the language model. This problem is well-known as \emph{hallucination}~\cite{ji2023survey-hallu} 
of generative language models. After augmenting the code snippet with the types of the other classes inferred by SnR, MLMTyper predicts the right type, \emph{com.google.gwt.dom.client.Document}, as the top-1 result. 

After this step, we obtain a list of candidate types predicted by MLMTyper for each API element in $ACS$ (or $CS$). Based on our observation, a considerable number of candidates are haphazardly generated. 
To solve this problem, we filter the candidates that do not exist in the pre-built KB of SnR. 
~The refined top-$k$ candidate types of an API element $e\in APIs(CS)$ are denoted as $types_{k}^{S}(e)$, where the superscript `$S$' stands for the statistically-based method. These candidates are subsequently used to reduce the pre-built KB by filtering the irrelevant types to improve the performance of SnR.

\subsection{Knowledge Base Reduction}\label{chap:knowledge base reduction}

Constraint-based type inference methods search for the types of API elements in a code snippet from the pre-built KB. Generally, the KB should cover a large number of API libraries to ensure a good capability of the methods. As a result, there can be multiple candidate types of an API element in the KB. If a code snippet does not contain sufficient constraints to determine the right type of an API element from the candidates, the methods will not be able to produce the desired answer. This limitation can be alleviated by leveraging the type inference results of statistically-based methods. Specifically, we introduce a KB reduction module in \MethodName{}.

The KB reduction module is proposed based on the assumption that the top-$k$ candidate types predicted by the statistically-based method for an API element $e$, i.e., $types_{k}^{S}(e)$, probably contain the right type of $e$. This assumption is supported by the evaluation results in~\cite{DBLP:conf/kbse/HuangYXX0022-prompt}, which reveal that the accuracy of the top-3 candidates predicted by MLMTyper is 91\% on a widely used dataset, StatType-SO~\cite{DBLP:conf/icse/PhanNTTNN18-statype}. Thus, the noise types of API elements in the pre-built KB of the constraint-based method can be reduced based on the top-$k$ candidates. Considering that the syntax knowledge of API elements is necessary for the constraint-based methods to solve the constraints extracted from code snippets, we implement the KB reduction module in two phases.

\textbf{Phase 1: Candidate Class \& Interface Type Collection.} In this phase, we build a set of candidate types, $CanCITypes$, that can probably contain the right types of all classes and interfaces in the input code snippet. The reason why we focus on the classes and interfaces is explained in Section~\ref{subsec:titd}. At first, we collect the top-$k$ candidate types predicted for each class or interface, resulting in an initial $CanCITypes$. Since the performance of existing statistically-based methods is not perfect (e.g., the precision and recall of MLMTyper are not 100\%), it is possible that the top-$k$ candidate types of a class or an interface do not contain the right type. Based on this consideration, if a class or an interface has a type inferred by the constraint-based method, and the type is not in $CanCITypes$, we add the type to $CanCITypes$ as it might be the right type. Finally, we formally define $CanCITypes$ as
\begin{equation}
\setlength{\abovedisplayskip}{0pt}
    CanCITypes = \left(\bigcup_{e \in CIs(CS)} types_{k}^{S}(e)\right) \cup \left\{type^{C}(e)|e \in Typed\_APIs^{C}(CS) \cap CIs(CS)\right\}
     \tiny
\end{equation}
where $CIs(CS)$ represents the set of classes and interfaces in $CS$. It holds that $CIs(CS)\subseteq APIs(CS)$.

\textbf{Phase 2: Reduced KB Construction.} In this phase, we construct a reduced KB by extracting the syntax knowledge of each class or interface type $ctype\in CanCITypes$ from the pre-built KB, denoted as $KB$, of the constraint-based method. The syntax knowledge includes the methods and fields declared in $ctype$, the super classes or interfaces of $ctype$ as well as their declared methods and fields. The reduced KB is formally defined as
\begin{equation}
    Reduced\_KB = \bigcup_{ctype \in CanCITypes} Syn\_Knowl(ctype, KB)
\end{equation}
where $Syn\_Knowl(ctype, KB)$ represents the syntax knowledge of $ctype$ in $KB$, i.e.,
\begin{equation}
    \tiny
    Syn\_Knowl(ctype, KB) = \bigcup_{e \in \{ctype\} \cup SupCITypes(ctype, KB)} e \cup e.Methods \cup e.Fields 
\end{equation}
where $SupCITypes(ctype, KB)$ represents the set of super class or interface types of $ctype$ in $KB$. For a class or an interface type, $e$, $e.Methods$ and $e.Fields$ are the set of methods and the set of fields declared in $e$, respectively. The reduced KB also preserves the relationships between the elements in it.

\subsection{Iterative Type Inference} 
\begin{figure}
    \centering
    \includegraphics[width=\linewidth]{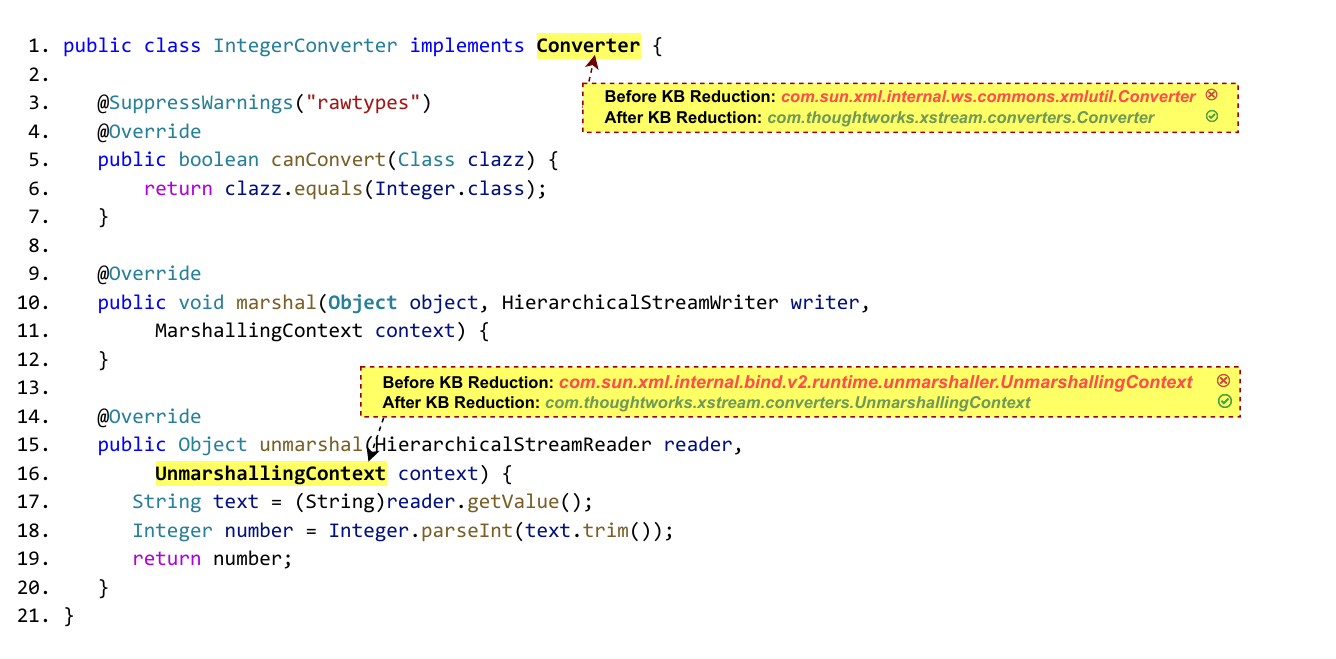}
    \caption{Example of the type inference results improved by leveraging the reduced knowledge base (KB) using a code snippet from the SO post `39005622'. 
    The types in the yellow background boxes are inferred by SnR before and after reducing the KB based on the top-$k$(=3) candidate types predicted by MLMTyper.}
    \label{fig:kb_reduction}
\end{figure}

Based on the reduced KB, we re-apply the constraint-based method to the input code snippet to obtain new inferred types of API elements. Since many noise types are excluded in the reduced KB, the quality of the inferred types could be improved. Fig.~\ref{fig:kb_reduction} shows a code snippet from the SO post `39005622'. Before reducing the KB, the types inferred by SnR for the interface \emph{Converter}[1,1] and the class \emph{UnmarshallingContext}[16,1] are both wrong. 
After reducing the KB, SnR infers the right types for them because the two wrong types are excluded in the reduced KB.
These cases demonstrate the benefit of our KB reduction strategy.

Furthermore, the new set of types with higher quality inferred by the constraint-based method can be used to better augment the code snippet and thus improve the performance of the statistically-based method again. Considering this \textbf{mutual promotion effect} between both methods, \MethodName{} adopts an iterative mechanism. 
To control the iteration, we set two conditions: 1) The inference results become stable or 2) The number of iterations exceeds a threshold $\delta$ (=10 in this work). The second condition is used to avoid infinite iteration due to repeated oscillations in the results. After performing a round of both methods, \MethodName{} examines the termination conditions. 
The iteration ends when either of the two conditions is satisfied.

\subsection{Type Inference Result Combination}\label{subsec:tirc}
After the iterative type inference process is terminated, we combine the inference results of the constraint- and statistically-based methods to produce the final inference results. The combination is performed in two phases.

\textbf{Phase 1: Combining the Inference Results of Each Method.} Through our observation from the experiments, there can be fluctuations in the results produced by either of the two methods during the iteration process. One case is that the inferred type(s) of an API element in one round may differ from those in another round. Another case is that the inference of an API element may fail in one round but succeed in another round. 
Unfortunately, there is no way to design the optimal combination strategy which can find the maximum number of right types inferred for the API elements from the results produced in different rounds, without knowing the ground truth. Here, we combine the inference results of each method using a simple strategy, i.e., keeping the last successfully inferred type(s) of each API element. After this phase, we denote the type of an API element, $e$, inferred by the constraint-based method as $comb\_type^{C}(e)$ and the top-$k$ candidate types of $e$ predicted by the statistically-based method as $comb\_types_{k}^{S}(e)$.

\textbf{Phase 2: Combining the Inference Results of Both Methods.} In this phase, we combine the combined results of both methods. Since the precision of constraint-based methods, e.g., SnR, is generally higher than that of statistically-based methods, e.g., 
MLMTyper (see Table~\ref{tab:rq2-component}), 
we combine the type(s) of an API element, $e$, inferred by both methods according to two criteria: 1) If only the statistically-based method predicts the top-$k$ candidate types of $e$, i.e., $comb\_type^{C}(e) == null \wedge comb\_types_{k}^{S}(e) \neq \emptyset$, then the top-1 type, i.e., $comb\_types_{1}^{S}(e)$, is determined as the type of $e$; 2) If both methods produce types for $e$, i.e., $comb\_type^{C}(e) \neq null \wedge comb\_types_{k}^{S}(e) \neq \emptyset$, then the type inferred by the constraint-based method, i.e., $comb\_type^{C}(e)$, is accepted as the type of $e$.

\section{Evaluation}
In this section, we evaluate \MethodName{} by answering the following three research questions:

\textbf{RQ1:} How do the internal settings affect the performance of \MethodName{}?
    
\textbf{RQ2:} Can \MethodName{} improve SOTA type inference methods?
    
\textbf{RQ3:} What are the contributions of the code context augmentation and KB reduction mechanisms used in \MethodName{}?

Our experimental environment is a workstation with an Intel(R) Xeon(R) Silver 4210R CPU (@2.40GHz) and an NVIDIA GeForce RTX 3090 GPU, running Ubuntu 20.04.3 LTS.


\subsection{Experimental Setup}

\subsubsection{Dataset}
We used two open-source datasets of incomplete Java code snippets collected from SO: StatType-SO~\cite{DBLP:conf/icse/PhanNTTNN18-statype} and Short-SO~\cite{DBLP:conf/kbse/HuangYXX0022-prompt}. StatType-SO contains 268 code snippets. 
The API elements included in each code snippet are primarily from one of six popular libraries, namely Android, GWT, Hibernate, JDK, Joda Time, and XStream. StatType-SO has been widely used in prior studies~\cite{DBLP:conf/icse/PhanNTTNN18-statype,DBLP:journals/scp/VelazquezRodriguezNR23-resico, DBLP:conf/kbse/HuangYXX0022-prompt,DBLP:conf/kbse/SaifullahAR19-coster,DBLP:conf/icse/DongGTS22}. 
Short-SO contains 120 code snippets.
The API elements in each code snippet are also related to one of the same six libraries as StatType-SO. The main difference between the two datasets is that all the code snippets in Short-SO contain less than three lines of code, whereas the code snippets in StatType-SO are longer with an average of 28 lines of code. 


\begin{table}
    \centering
    \caption{Average precision and recall (which are the same) of \MethodName{} with different settings of top-$k$.}
    \label{tab:rq1-topk-iteration}
    \resizebox{\linewidth}{!}{
\begin{tabular}{cccccccccc} 
\hline
\multirow{2}{*}{Iteration} & \multicolumn{4}{c}{StatType-SO}                                           &  & \multicolumn{4}{c}{Short-SO}                                               \\ 
\cline{2-5}\cline{7-10}
                           & Top-1            & Top-3            & Top-5            & Top-10           &  & Top-1            & Top-3            & Top-5            & Top-10            \\ 
\hline
1                          & 95.77\%          & 95.68\%          & 95.73\%          & 95.77\%          &  & 90.86\%          & 90.86\%          & 90.86\%          & 90.86\%           \\
2                          & \textbf{97.53\%} & \textbf{97.53\%} & \textbf{97.53\%} & \textbf{97.53\%} &  & \textbf{92.52\%} & \textbf{92.52\%} & \textbf{92.52\%} & \textbf{92.52\%}  \\
3                          & 97.37\%          & 97.37\%          & 97.37\%          & 97.37\%          &  & 92.52\%          & 92.52\%          & 92.52\%          & 92.52\%           \\
4                          & 97.31\%          & 97.31\%          & 97.31\%          & 97.31\%          &  & 92.52\%          & 92.52\%          & 92.52\%          & 92.52\%           \\
5                          & 97.31\%          & 97.31\%          & 97.31\%          & 97.31\%          &  & 92.52\%          & 92.52\%          & 92.52\%          & 92.52\%           \\
\hline
\end{tabular}
    }
\end{table}

\subsubsection{Implementation of Baselines and \MethodName{}}\label{sec:baselines}
As described in Section~\ref{sec:meth}, we implemented \MethodName{} by integrating two SOTA methods: the constraint-based method, SnR~\cite{DBLP:conf/icse/DongGTS22}, and the statistically-based method, MLMTyper~\cite{DBLP:conf/kbse/HuangYXX0022-prompt}. Therefore, SnR and MLMTyper are used as our baselines. We implemented them based on the replication package\footnote{\url{https://doi.org/10.5281/zenodo.5843327}} of SnR released at Zenodo and the replication package\footnote{\url{https://github.com/SE-qinghuang/ASE-22-TypeInference}} of MLMTyper released at GitHub. In addition, since ChatGPT~\cite{ChatGPT} has been widely used by developers to address various software engineering tasks, we also chose it as a baseline. We used the GPT-3.5 (text-davinci-002-render-sha) model for the evaluation of ChatGPT, which is available on the webpage\footnote{\url{https://chat.openai.com}} provided by OpenAI. 

\subsubsection{Metrics}\label{subsec:metrics}
We used two widely used metrics: precision and recall. For a code snippet, precision measures the percentage of the types correctly inferred; and recall measures the percentage of the API elements whose types are correctly inferred. 
\begin{equation}
        Precision = \frac{\#\text{Correctly Inferred Types}}{\#\text{Inferred~Types}}    
\end{equation}
\begin{equation}
        Recall = \frac{\#\text{Correctly Inferred Types}}
        {\#\text{Requested Types}}   
\end{equation}
where `\#Requested Types' is the number of API elements whose types are requested to be inferred. 
\textbf{For the methods, e.g., MLMTyper, ChatGPT, and \MethodName{}, that can infer types for all requested API elements, their precision and recall are the same on a code snippet.}

Note that for a fair comparison, we only consider the top-1 type inferred for an API element by MLMTyper. For each of the two datasets, after measuring the precision and recall of a method on every code snippet, we calculated the average precision and recall of the method on the set of code snippets related to each of the six libraries as well as the average precision and recall of the method on all code snippets. Moreover, we used the Wilcoxon signed-rank test~\cite{wilcoxon1992individual-wilcoxonrank} to examine whether the performance of \MethodName{} is significantly better than that of the baselines and ChatGPT.

\subsection{RQ1: How do the internal settings affect the performance of \MethodName{}?}
     
\textbf{Motivation.} 
In \MethodName{}, there are two changeable settings. One is the parameter top-$k$, which indicates the number of candidate types predicted by the statistically-based method that are used to reduce the pre-built KB of the constraint-based method. The setting of $k$ may have impacts on the quality of the reduced KB and thus affect the performance of the constraint-based method. Intuitively, a large $k$ may retain too many noise types of API elements, while a small $k$ may exclude the right types of some API elements. The other one is the execution order of the constraint- and statistically-based methods. As shown in Fig.~\ref{fig:overview}, \MethodName{} applies the constraint-based method before the statistically-based method. However, it is able to change the execution order of both methods. We need to validate the impacts of $k$ and the execution order on the performance of \MethodName{}. Moreover, since \MethodName{} is an iterative framework, we want to investigate the number of iterations required by \MethodName{} to produce the inference results for code snippets, as well as the performance of \MethodName{} during the iterations. 

\textbf{Approach.} 
To validate the impact of $k$, we applied \MethodName{} to each code snippet in StatType-SO and Short-SO, with different settings of $k\in \{1, 3, 5, 10\}$. For each setting of $k$, we measured the average precision and recall of \MethodName{} on all code snippets in each dataset. We also measured the average precision and recall of \MethodName{} based on the intermediate results produced after each iteration to investigate the convergence of the iterative process and the performance change of \MethodName{} during the iterations. 
After determining the proper setting of $k$, we further validated the impacts of the execution order of the constraint- and statistically-based methods by implementing a variant of \MethodName{}. The variant exchanged the order of both methods, which is referred to as \MethodName{}-SC. We applied \MethodName{}-SC to all code snippets and measured the average precision and recall of \MethodName{} on each dataset. As mentioned previously, the precision and recall of \MethodName{} are always the same because it can infer types for all API elements in every code snippet.

\textbf{Results.} 
\textbf{The setting of $k$ has negligible impacts on the performance of \MethodName{}.} Table~\ref{tab:rq1-topk-iteration} presents the average precision and recall of \MethodName{} on the two datasets under different $k$ values within five iterations. The performance of \MethodName{} only has a little difference at the first iteration on StatType-SO and remains the same in the other cases. This result means that the setting of $k$ nearly has no impact on the performance of \MethodName{}. Through in-depth analysis, most of the candidate types predicted by MLMTyper are haphazardly generated by the language model. When constructing the top-$k$ candidate types for an API element, we filter the invalid candidates that do
not exist in the pre-built KB of SnR. As a result, the top-$k$ candidate types often contain only 0-2 types. Moreover, if MLMTyper could predict the right type of an API element, the right type is often the first in the top-$k$ candidates. This is the reason why the different settings of $k$ do not affect the performance of \MethodName{}. We set $k=3$ in the subsequent experiments to ensure generality.

\begin{figure}
    \centering
    \includegraphics[width=0.85\linewidth]{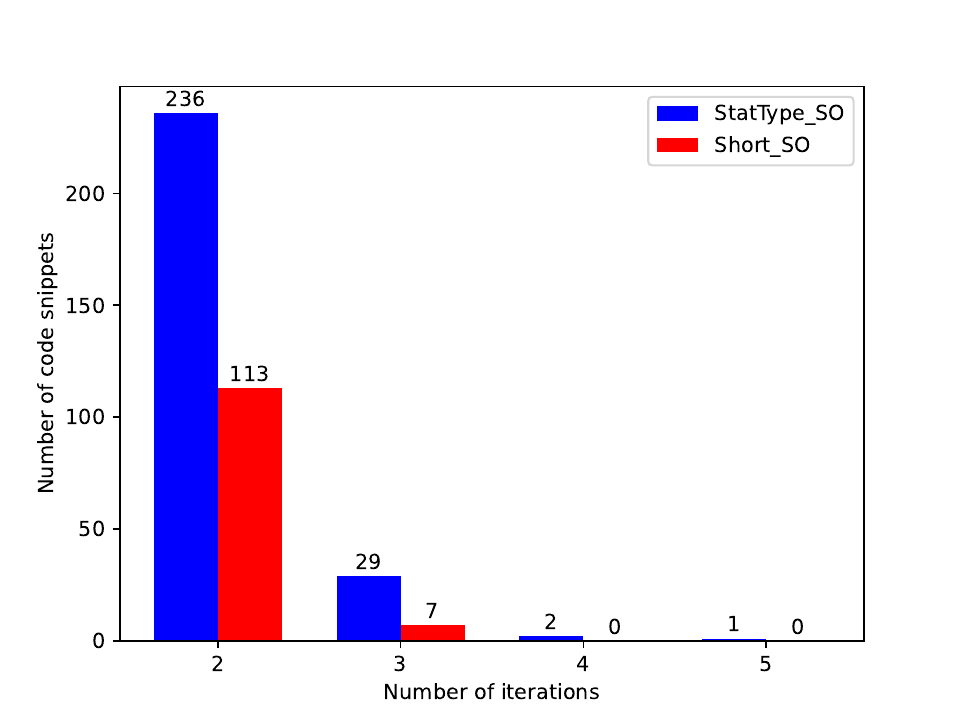}
    \caption{Distribution of the number of iterations of \MethodName{}.}
    \label{fig:rq1-iteration_number}
\end{figure}

\begin{table}
    \centering
    \caption{Average precision and recall (which are the same) of \MethodName{}-SC.} 
    \label{tab:variant}
    \resizebox{0.8\linewidth}{!}{
\begin{tabular}{cccccc} 
\hline
\multirow{2}{*}{Iteration} & \multicolumn{2}{c}{StatType-SO~ ~} &  & \multicolumn{2}{c}{Short-SO~ ~}  \\ 
\cline{2-6}
                           & Precision & Recall                 &  & Precision & Recall               \\ 
\hline
1                          & 86.62\%   & 86.62\%                &  & 84.29\%   & 84.29\%              \\
2                          & 86.77\%   & 86.77\%                &  & 84.50\%   & 84.50\%              \\
3                          & 86.71\%   & 86.71\%                &  & 84.50\%   & 84.50\%              \\
4                          & 86.77\%   & 86.77\%                &  & 84.50\%   & 84.50\%              \\
5                          & 86.71\%   & 86.71\%                &  & 84.50\%   & 84.50\%              \\
\hline
\end{tabular}
    }
\end{table}

\textbf{\MethodName{} converges within two or three iterations on 99.2\% code snippets.} Fig.~\ref{fig:rq1-iteration_number} shows the distribution of the number of iterations performed by \MethodName{} on the code snippets in two datasets. \MethodName{} converges on 349 (89.9\%), 36 (9.3\%), 2 (0.5\%), and 1 (0.3\%) code snippets after two, three, four, and five iterations, respectively.

\begin{table*}[htbp]
    \centering
    \caption{Average precision and recall of four methods. For MLMTyper, ChatGPT, and iJTyper, their respective precision and recall are the same.}
    \label{tab:rq2-baseline_compare}
    \resizebox{0.99\textwidth}{!}{
\begin{tabular}{cccccccccccc} 
\hline
\multirow{2}{*}{Library} & \multicolumn{5}{c}{StatType-SO}                                                                                                                                                                                                                                                                                                        &  & \multicolumn{5}{c}{Short-SO}                                                                                                                                                                                                                                                                                                            \\ 
\cline{2-6}\cline{8-12}
                         & \begin{tabular}[c]{@{}c@{}}MLMTyper \\Precision/Recall\end{tabular} & \begin{tabular}[c]{@{}c@{}}SnR \\Precision\end{tabular} & \begin{tabular}[c]{@{}c@{}}SnR \\Recall\end{tabular} & \begin{tabular}[c]{@{}c@{}}ChatGPT \\Precision/Recall\end{tabular} & \begin{tabular}[c]{@{}c@{}}iJTyper\\Precision/Recall\end{tabular} &  & \begin{tabular}[c]{@{}c@{}}MLMTyper \\Precision/Recall\end{tabular} & \begin{tabular}[c]{@{}c@{}}SnR \\Precision\end{tabular} & \begin{tabular}[c]{@{}c@{}}SnR \\Recall\end{tabular} & \begin{tabular}[c]{@{}c@{}}ChatGPT \\Precision/Recall\end{tabular} & \begin{tabular}[c]{@{}c@{}}iJTyper\\Precision/Recall\end{tabular}  \\ 
\hline
Android                  & 71.91\%                                                                & 98.36\%                                                 & 92.59\%                                              & 97.89\%                                                               & 97.22\%                                                              &  & 69.09\%                                                                & 88.68\%                                                 & 85.45\%                                              & 95.83\%                                                               & 89.09\%                                                               \\
GWT                      & 78.71\%                                                                & 97.48\%                                                 & 86.55\%                                              & 95.98\%                                                               & 98.89\%                                                              &  & 68.89\%                                                                & 87.80\%                                                 & 80.00\%                                              & 92.68\%                                                               & 86.67\%                                                               \\
Hibernate                & 42.30\%                                                                & 96.35\%                                                 & 95.08\%                                              & 88.70\%                                                               & 93.77\%                                                              &  & 38.60\%                                                                & 89.29\%                                                 & 87.72\%                                              & 70.37\%                                                               & 89.47\%                                                               \\
JDK                      & 86.87\%                                                                & 100.00\%                                                & 92.42\%                                              & 100.00\%                                                              & 100.00\%                                                             &  & 90.74\%                                                                & 98.04\%                                                 & 92.59\%                                              & 100.00\%                                                              & 98.15\%                                                               \\
Joda Time                & 59.80\%                                                                & 94.83\%                                                 & 82.91\%                                              & 98.83\%                                                               & 95.98\%                                                              &  & 66.67\%                                                                & 94.59\%                                                 & 72.92\%                                              & 100.00\%                                                              & 100.00\%                                                              \\
XStream                  & 55.78\%                                                                & 91.53\%                                                 & 90.44\%                                              & 82.96\%                                                               & 98.01\%                                                              &  & 56.47\%                                                                & 92.77\%                                                 & 90.59\%                                              & 93.15\%                                                               & 91.76\%                                                               \\ 
\hline
Average                  & 65.90\%***                                                             & 96.43\%**                                               & 90.00\%***                                           & 94.06\%*                                                              & 97.31\%                                                              &  & 65.08\%***                                                             & 91.86\%*                                                & 84.88\%***                                           & 92.01\%                                                               & 92.52\%                                                               \\ 
\hline
\multicolumn{12}{c}{***p<0.001, **p<0.01, *p<0.05}                                                                                                                                                                                                                                                                                                                                                                                                                                                                                                                                                                                                                                                               
\end{tabular}
        }
\end{table*}

\textbf{\MethodName{} achieves the best performance after the second iteration.} 
From the results in Table~\ref{tab:rq1-topk-iteration}, \MethodName{} achieves high precision and recall of over 90\% after the first iteration. After the second iteration, \MethodName{} reaches the best performance, with a precision and recall of 97.53\% and 92.52\% on the two datasets, respectively. This result confirms the necessity of the iterative type inference mechanism used in \MethodName{}. 
However, with more iterations, the performance of \MethodName{} decreases slightly and finally becomes stable on StatType-SO. Through in-depth analysis, the issue is due to the type inference for two classes, \textit{javax.persistence.Entity} and \textit{javax.persistence.Table}, in a code snippet. \MethodName{} correctly inferred them in the second iteration. After that, the former was inferred as \textit{org.hibernate.annotations.Entity} in the third iteration, and the latter was inferred as \textit{org.hibernate.annotations.Table} in the fourth iteration. The inference results of the fifth iteration are the same as those of the fourth iteration, and thus the iterative process converged. In fact, both \textit{org.hibernate.annotations.Entity/Table} and \textit{javax.persistence.Entity/Table} can be used in the context of the code snippet, but the latter is more portable and recommended.

\textbf{Exchanging the execution order of the constraint- and statistically-based methods in \MethodName{} leads to lower performance.} Table~\ref{tab:variant} shows the average precision and recall of \MethodName{}-SC on the two datasets. We observe a performance oscillation between 86.77\% and 86.71\% on StatType-SO, which is caused by the inference of a class switched back and forth between the right type, \textit{org.hibernate.cfg.Configuration}, and a wrong type, \textit{java.util.Configur\newline ation}. Moreover, the performance of \MethodName{}-SC on both datasets is much worse than that of \MethodName{}, i.e., 86.71\% versus 97.31\% on StatType-SO and 84.50\% versus 92.52\% on Short-SO. We tracked the inference process and found that MLMTyper failed to predict the right types for a considerable number of API elements at the beginning. The low precision of MLMTyper has negative impacts on the KB reduction and thus affects the performance of SnR and iJTyper.

\subsection{RQ2: Can \MethodName{} improve SOTA type inference methods?}

\textbf{Motivation.} 
\MethodName{} aims to infer types of API elements in Java code snippets. It is necessary to evaluate the effectiveness of \MethodName{} by comparing it with SOTA methods. Moreover, we want to examine whether \MethodName{} can achieve comparable or better results than the popular and widely used model, ChatGPT.

\begin{figure*} 
        \centering
        \includegraphics[width=0.99\textwidth]{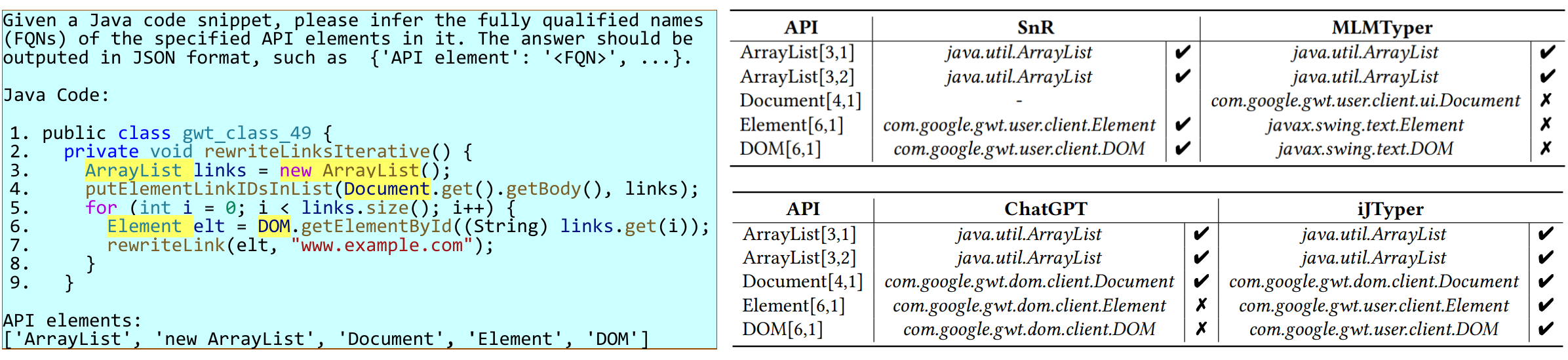}
        \caption{Example of evaluating ChatGPT in type inference.}
        \label{fig:rq2-chatgpt-prompt}
\end{figure*}

\textbf{Approach.} 
We used two SOTA type inference methods, i.e., SnR and MLMTyper, as baselines. We implemented them based on their replication packages and applied them to the code snippets in the two datasets. For ChatGPT, we used its implementation based on the GPT-3.5 model. A prompt template shown in Fig.~\ref{fig:rq2-chatgpt-prompt} was used to ask ChatGPT for the FQNs (i.e., types) of the API elements that we focus on in each code snippet from both datasets. After that, we measured the average precision and recall of SnR, MLMTyper, ChatGPT, and \MethodName{} on the code snippets related to each of the six libraries. We also measured the average precision and recall of the methods on all code snippets in each dataset.
    
\textbf{Results.} 
\textbf{\MethodName{} significantly outperforms SnR and MLMTyper in terms of both precision and recall.} Table~\ref{tab:rq2-baseline_compare} presents the results of \MethodName{} and three baselines. \MethodName{} achieves the highest average precision and recall, i.e., 97.31\% and 92.52\%, on StatType-SO and Short-SO, respectively. The precision/recall of MLMTyper is much lower than that (i.e., 91\%) of MLMTyper reported in~\cite{DBLP:conf/kbse/HuangYXX0022-prompt}. This is because although MLMTyper can predict the right types for most API elements in the top-3 candidate types, the right types do not always appear in the top-1. \MethodName{} outperforms SnR with respect to almost all the six libraries on both datasets. For example, SnR often makes mistakes with \textit{java.util.Date} and \textit{java.sql.Date}, while \MethodName{} can distinguish between these types. \MethodName{} significantly improves the recall of SnR and MLMTyper by at least 7.31\% and 27.44\%, respectively. In terms of precision, \MethodName{} significantly improves SnR by 0.88\% on StatType-SO and by 0.66\% on Short-SO. 

\textbf{\MethodName{} significantly improves ChatGPT by 3.25\% in terms of recall on StatType-SO.} On Short-SO, \MethodName{} improves the average recall of ChatGPT by 0.51\%, however the improvement is not significant. \MethodName{} outperforms ChatGPT on the code snippets related to GWT, Hibernate, and XStream in StatType-SO, and is slightly worse than ChatGPT on the code snippets related to Android and Joda Time. 
For Hibernate, ChatGPT often mistakenly infers the types under \textit{org.hibernate} as those under \textit{javax}, e.g., \textit{org.hibernate.validator.AssertTrue} and \textit{javax.validation.constraints. AssertTrue}. The former is a general standard, whereas the latter is a specific implementation under Hibernate. ChatGPT may not be able to accurately determine which package is expected, so it chooses the common and general one in its training corpus. For XStream, ChatGPT often incorrectly infers the types under \textit{com.thoughtworks.xstream.converters}, e.g., mistakenly inferring \textit{com. thoughtworks.xstr} as \textit{com.thoughtworks.xstream.core.MarshallingCon\newline text}. A possible explanation is that there is not enough information about this package in the training corpus to make ChatGPT remember the types under it. \MethodName{} performs better in these situations because of its fusion of knowledge.



\subsection{RQ3: What are the contributions of the code context augmentation and KB reduction mechanisms used in \MethodName{}?}
\textbf{Motivation.} 
There are two key mechanisms adopted in \MethodName{}: 1) Code context augmentation which is used to enhance the context of the input code snippet for the statistically-based method; and 2) KB reduction which is used to assist the constraint-based method in filtering irrelevant types. We need to validate whether these mechanisms could improve the corresponding methods.

\textbf{Approach.} We measured the precision and recall of the two type inference methods inside \MethodName{} separately to study the promotion effects of the two mechanisms. Note that although the two internal methods originate from SnR and MLMTyper, their inference results are different from the original methods because they are improved by iteratively utilizing the two mechanisms. We referred to the internal constraint- and statistically-based methods as \MethodName{}-C and \MethodName{}-S, respectively. For each of these two methods, we obtained the final successfully inferred results on a code snippet during the iterative process. We measured the average precision and recall of both methods for each library and each dataset. 
\begin{table}
    \centering
    \caption{Average precision of 
    \MethodName{}-C and \MethodName{}-S, in comparison with the original SnR and MLMTyper.}
    \label{tab:rq2-component}
    \resizebox{\linewidth}{!}
        {
            \begin{tabular}{cccccccccc} 
    \hline
    \multirow{2}{*}{Library} & \multicolumn{4}{c}{StatType-SO}      &  & \multicolumn{4}{c}{Short-SO}        \\ 
    \cline{2-5}\cline{7-10}
                             & MLMTyper   & iJTyper-S & SnR       & iJTyper-C &  & MLMTyper   & iJTyper-S & SnR     & iJTyper-C  \\ 
    \hline
    Android                  & 71.91\%    & 82.41\%   & 98.36\%   & 99.02\%   &  & 69.09\%    & 85.45\%   & 88.68\% & 90.57\%    \\
    GWT                      & 78.71\%    & 93.04\%   & 97.48\%   & 98.81\%   &  & 68.89\%    & 75.56\%   & 87.80\% & 90.48\%    \\
    Hibernate                & 42.30\%    & 61.64\%   & 96.35\%   & 94.35\%   &  & 38.60\%    & 71.93\%   & 89.29\% & 91.07\%    \\
    JDK                      & 86.87\%    & 99.49\%   & 100.00\%  & 100.00\%  &  & 90.74\%    & 98.15\%   & 98.04\% & 98.04\%    \\
    Joda Time                & 59.80\%    & 94.47\%   & 94.83\%   & 97.18\%   &  & 66.67\%    & 97.92\%   & 94.59\% & 100.00\%   \\
    XStream                  & 55.78\%    & 86.85\%   & 91.53\%   & 98.40\%   &  & 56.47\%    & 87.06\%   & 92.77\% & 92.77\%    \\ 
    \hline
    Average                  & 65.90\%*** & 86.32\%   & 96.43\%** & 97.96\%   &  & 65.08\%*** & 86.01\%   & 91.86\% & 93.82\%    \\ 
    \hline
    \multicolumn{10}{c}{***p<0.001, **p<0.01, *p<0.05}                                
    \end{tabular}
        }
\end{table}

\textbf{Results.} 
\textbf{The code context augmentation and KB reduction mechanisms significantly promote the constraint- and statistically-based methods.} Table~\ref{tab:rq2-component} and Table~\ref{tab:rq2-component-recall} present the average precision and recall respectively of \MethodName{}-C and \MethodName{}-S as well as SnR and MLMTyper. \MethodName{}-C increases the precision of SnR by 1.53\% on StatType-SO and by 1.96\% on Short-SO. In terms of recall, \MethodName{}-C improves SnR by 2.65\% and 2.03\% on both datasets. Most of the improvements are significant except for the precision on Short-SO. Similarly, \MethodName{}-S significantly improves MLMTyper by 20.42\% on StatType-SO and by 20.93\% on Short-SO in terms of both precision and recall. All these results demonstrate that the code context augmentation and KB reduction mechanisms used in \MethodName{} can promote the performance of the constraint- and statistically-based methods. 


\begin{table}
    \centering
    \caption{Average recall of 
    \MethodName{}-C and \MethodName{}-S, in comparison with the original SnR and MLMTyper.}
    \label{tab:rq2-component-recall}
    \resizebox{\linewidth}{!}
        {
           \begin{tabular}{cccccccccc} 
    \hline
    \multirow{2}{*}{Library} & \multicolumn{4}{c}{StatType-SO}          &  & \multicolumn{4}{c}{Short-SO}            \\ 
    \cline{2-5}\cline{7-10}
                             & MLMTyper   & iJTyper-S & SnR        & iJTyper-C &  & MLMTyper   & iJTyper-S & SnR      & iJTyper-C  \\ 
    \hline
    Android                  & 71.91\%    & 82.41\%   & 92.59\%    & 93.21\%   &  & 69.09\%    & 85.45\%   & 85.45\%  & 87.27\%    \\
    GWT                      & 78.71\%    & 93.04\%   & 86.55\%    & 92.20\%   &  & 68.89\%    & 75.56\%   & 80.00\%  & 84.44\%    \\
    Hibernate                & 42.30\%    & 61.64\%   & 95.08\%    & 93.11\%   &  & 38.60\%    & 71.93\%   & 87.72\%  & 89.47\%    \\
    JDK                      & 86.87\%    & 99.49\%   & 92.42\%    & 92.93\%   &  & 90.74\%    & 98.15\%   & 92.59\%  & 92.59\%    \\
    Joda Time                & 59.80\%    & 94.47\%   & 82.91\%    & 86.43\%   &  & 66.67\%    & 97.92\%   & 72.92\%  & 77.08\%    \\
    XStream                  & 55.78\%    & 86.85\%   & 90.44\%    & 98.01\%   &  & 56.47\%    & 87.06\%   & 90.59\%  & 90.59\%    \\ 
    \hline
    Average                  & 65.90\%*** & 86.32\%   & 90.00\%*** & 92.65\%   &  & 65.08\%*** & 86.01\%   & 84.88\%* & 86.91\%    \\ 
    \hline
    \multicolumn{10}{c}{***p<0.001, **p<0.01, *p<0.05}                                                                               
    \end{tabular}
        }
\end{table}

\section{Discussion}

\textbf{Practicality.} Type inference of API elements in incomplete code snippets is necessary to work with the code snippets. 
As demonstrated in our evaluations, \MethodName{} achieves high precision and recall and thus can effectively support a wide range of tasks, such as API usage mining~\cite{zhang2018code-usagemining,uddin2020mining-scene}, code search~\cite{thummalapenta2007parseweb,maji2021dcom}, and code repair~\cite{DBLP:conf/issta/TerragniLC16-csnippex,mesbah2019deepdelta}. 

\textbf{Limitation.} \MethodName{} iteratively integrates two methods, which has limitations in time efficiency. We performed \MethodName{} on the two datasets for three times and measured the average time cost of three main components spent on a code snippet in each dataset. Table~\ref{tab:time-efficiency} presents the results. 
On average, \MethodName{} takes about 104.02 and 51.33 seconds to analyze a code snippet in StatType-SO and Short-SO, respectively. \MethodName{} is more suitable for application scenarios that require high quality of inference results but do not require high efficiency. It is also worth noticing that \MethodName{} can achieve relatively high recall after the first iteration, allowing users to trade off between performance and efficiency. Moreover, although \MethodName{} combines the strengths of constraint- and statistically-based methods, it still cannot overcome some problems inherent in both methods, e.g., the out-of-vocabulary issue.

\textbf{Threats to Validity.} 
Threats to internal validity relate to the errors in the implementation of \MethodName{} and the baselines. 
We implemented the baselines, SnR and MLMTyper, based on their replication packages. For \MethodName{}, we double-checked the implementation code and tracked the inputs, intermediate results, and outputs of some code snippets to ensure correctness. 

Threats to external validity relate to the generalizability of the results. We conducted the experiments on two open-source datasets of incomplete Java code snippets related to six popular libraries. Both datasets are widely used in prior studies~\cite{DBLP:conf/icse/PhanNTTNN18-statype,DBLP:journals/scp/VelazquezRodriguezNR23-resico, DBLP:conf/kbse/HuangYXX0022-prompt,DBLP:conf/kbse/SaifullahAR19-coster,DBLP:conf/icse/DongGTS22}. The code snippets in both datasets are characterized by different lengths, and the six libraries have different applications. Based on these features, our evaluation results could have a good generalizability.

\begin{table}
    \centering
    \caption{Average time cost (in seconds) of \MethodName{} and its three main components.}
    \label{tab:time-efficiency}
    \resizebox{\linewidth}{!}
        {
            \begin{tabular}{ccccclcccc} 
\hline
\multirow{2}{*}{Library} & \multicolumn{4}{c}{StatType-SO}                                                      & \multicolumn{1}{c}{} & \multicolumn{4}{c}{Short-SO}                                                         \\ 
\cline{2-5}\cline{7-10}
                         & SnR    & MLMTyper & \begin{tabular}[c]{@{}c@{}}KB \\Reduction\end{tabular} & iJTyper &                      & SnR    & MLMTyper & \begin{tabular}[c]{@{}c@{}}KB\\Reduction\end{tabular} & iJTyper  \\ 
\hline
Android                  & 49.84~ & 29.12~   & 5.00~                                                  & 84.87~  &                      & 25.14~ & 14.67~   & 2.08~                                                 & 44.24~   \\
GWT                      & 75.89~ & 48.58~   & 6.03~                                                  & 134.87~ &                      & 28.41~ & 14.63~   & 2.15~                                                 & 46.33~   \\
Hibernate                & 62.09~ & 36.01~   & 5.09                                                   & 106.34~ &                      & 24.31~ & 16.09~   & 2.20~                                                 & 48.01~   \\
JDK                      & 71.08~ & 44.06~   & 10.85~                                                 & 125.83~ &                      & 23.15~ & 14.78~   & 2.45~                                                 & 41.19~   \\
Joda Time                & 46.22~ & 25.03~   & 2.60~                                                  & 88.64~  &                      & 30.10~ & 17.32~   & 2.15~                                                 & 62.25~   \\
XStream                  & 44.16~ & 27.87~   & 4.09~                                                  & 83.57~  &                      & 35.20~ & 19.90~   & 3.91~                                                 & 65.98~   \\ 
\hline
Average                  & 58.21~ & 35.11~   & 5.61~                                                  & 104.02~ &                      & 27.72~ & 16.23~   & 2.49~                                                 & 51.33~   \\
\hline
\end{tabular}
        }
\end{table}

\section{Related Work}
Prior studies on type inference for incomplete Java code snippets can be broadly categorized as constraint-based or statistical-based. These two kinds of methods have complementary characteristics. Our \MethodName{} combines them to exploit their respective advantages. Moreover, there are a number of studies on type inference for dynamically typed languages, e.g., JavaScript and Python.

\textbf{Constraint-based Type Inference.} Baker~\cite{DBLP:conf/icse/SubramanianIH14-baker} maintains a candidate type list for each API and tries to reduce the candidates based on the constraints in the code snippet using a set of heuristic rules. This procedure is repeated until all APIs are resolved or the approach is unable to shorten the candidate lists anymore. 
DepRes~\cite{DBLP:conf/kbse/Shokri21-ase_depres} generates a sketch for each API and uses a Z3 SMT solver~\cite{DBLP:conf/tacas/MouraB08-Z3} to handle the constraints.
SnR~\cite{DBLP:conf/icse/DongGTS22} goes one step further than them in handling generic types. 
All these methods pre-build a KB from Jar files and require a partial AST for type inference. However, the generation of partial ASTs is limited by the completeness of code snippets, which may lead to low recall in some cases.

\textbf{Statistically-based Type Inference.} StatType~\cite{DBLP:conf/icse/PhanNTTNN18-statype} considers both type context and resolution context. It solves type inference tasks using a machine translation model. COSTER~\cite{DBLP:conf/kbse/SaifullahAR19-coster} calculates a likelihood score, a context similarity score, and a name similarity score based on an occurrence likelihood dictionary to infer the type of an API. The dictionary is built from a code base. RESICO~\cite{DBLP:journals/scp/VelazquezRodriguezNR23-resico} views type inference as a text classification task and uses a classifier to infer the type of an API along with its context. Huang et al.~\cite{DBLP:conf/kbse/HuangYXX0022-prompt} transforms the type inference problem into a fill-in-blank task and aligns it with the prompt-tuning of a large pre-trained code model. These methods are limited by the training corpus. When encountering the APIs and code contexts that are less frequent in the training corpus, the precision may be low. 

\textbf{Type Inference for Dynamically-typed Languages.} JSNICE~\cite{DBLP:conf/popl/RaychevVK15-JSNICE} predicts names of identifiers and type annotations of variables for JavaScript using conditional random field. NL2Type~\cite{DBLP:conf/icse/MalikPP19-NL2Type} utilizes natural language information in source code, including annotations, function names, and parameter names, to predict type signatures of JavaScript functions. It builds a recurrent LSTM network and formulates type inference as a classification problem. LambdaNet~\cite{DBLP:conf/iclr/WeiGDD20-lambda_net} generates a type dependency graph (TDG) for TypeScript source code and then uses a graph neural network to infer types. HiTyper~\cite{DBLP:conf/icse/PengGLG0ZL22-Hityper} hybrids deep learning-based and static type inference. It obtains type recommendations from a neural network and performs type rejection and static inference on a TDG built from Python source code. HiTyper differs from \MethodName{} in that it only mixes the two methods without enhancing their individual performance.
~Moreover, these methods infer types for source code written in dynamically-typed languages, which should be complete and syntactically correct. In contrast, our \MethodName{} focuses on incomplete Java code snippets. 

\section{Conclusion}
We propose \MethodName{}, an iterative type inference framework for Java code snippets by integrating a constraint-based method and a statistically-based method. \MethodName{} executes the two methods iteratively and performs code context augmentation as well as KB reduction to improve the performance. Evaluation results on two open-source datasets demonstrate that \MethodName{} successfully combines the advantages of both methods and achieves the highest precision/recall of 97.31\% and 92.52\% on both datasets. \MethodName{} significantly improves the state-of-the-art baselines, SnR and MLMTyper, by at least 7.31\% and 27.44\% in terms of recall, respectively. \MethodName{} is flexible and can be extended for integrating other constraint- and statistically-based methods. 
In future work, we plan to apply \MethodName{} to some software engineering tasks, such as API usage mining and code repair. 

\section{Data Availability}
We release the replication package of this work at GitHub\footnote{\url{https://anonymous.4open.science/r/iJTyper-0A4D}} to help other researchers reproduce and extend our study.



\bibliographystyle{ACM-Reference-Format}
\bibliography{refs}



\end{document}